\def\mK{{\rm \mu K}}

\def\expec#1{\langle#1\rangle}

\def\etal{{\frenchspacing\it et al.}}
\def\ie{{\frenchspacing\it i.e.}}
\def\eg{{\frenchspacing\it e.g.}}
\def\etc{{\frenchspacing\it etc.}}
\def\crr{\cr\noalign{\vskip 4pt}}
\def\pp{\noindent\parshape 2 0truecm 13.6truecm 1truecm 12.6truecm}
\def\rf#1;#2;#3;#4 {\par\pp#1, {\it #2}, {\bf #3}, #4. \par}
\def\rg#1;#2;#3;#4;#5 {\par\pp#1, {\it #2}, {\bf #3}, #4 (``#5"). \par}

\def\beq#1{\begin{equation}\label{#1}}
\def\eeq{\end{equation}}
\def\beqa#1{\begin{eqnarray}\label{#1}}
\def\eeqa{\end{eqnarray}}
\def\eq#1{equation~(\ref{#1})}
\def\Eq#1{Equation~(\ref{#1})}
\def\eqnum#1{~(\ref{#1})}

\def\bfig{\begin{figure}[h] \centerline{\hbox{}}\vfill}
\def\efig{\end{figure}\vfill\newpage}

\def\fig#1{Figure~\ref{#1}}
\def\Fig#1{Figure~\ref{#1}}
\def\fignum#1{~\ref{#1}}

\def\sec#1{Section~\ref{#1}}
\def\secnum#1{~\ref{#1}}

\def\ns{\vskip-0.2truecm}
\def\ns{}

\def\spose#1{\hbox to 0pt{#1\hss}}
\def\simlt{\mathrel{\spose{\lower 3pt\hbox{$\mathchar"218$}}
     \raise 2.0pt\hbox{$\mathchar"13C$}}}
\def\simgt{\mathrel{\spose{\lower 3pt\hbox{$\mathchar"218$}}
     \raise 2.0pt\hbox{$\mathchar"13E$}}}
\def\simpropto{\mathrel{\spose{\lower 3pt\hbox{$\mathchar"218$}}
     \raise 2.0pt\hbox{$\propto$}}}

\def\addr#1{{\small\it #1}}
\def\auth#1{{#1}}

\def\ed{\end{document}}

\def\vx{{\bf x}}
\def\vk{{\bf k}}

\def\va{{\bf a}}
\def\n{\varepsilon}
\def\vn{{\bf \n}}

\def\N{N}

\def\scale{\gamma}

\def\Qrmsps{Q_{rms-ps}}

\def\l{\ell}
\def\summ{\sum_{m=-\l}^{\l}}
\def\summp{\sum_{m=-\l}^{\l}}

\def\sumlp{\sum_{\l=2}^{\infty}}
\def\sumk{\sum_{k=1}^{N''}}
\def\expec#1{\langle#1\rangle}
\def\tr{\hbox{tr}\>}
\def\Dl{\Delta\l}

\def\D{D}
\def\Dest{\tilde D}
\def\Dt{\Dest}
\def\vz{{\bf z}}
\def\ve{{\bf e}}
\def\vek{{\ve_k}}
\def\vekp{{\ve_{k'}}}
\def\veone{{\ve_1}}
\def\weight{\alpha}

\def\Ell{{\cal L}}
\def\E{E}
\def\El{\E}
\def\V{V}
\def\v{v}
\def\u{p}
\def\bias{B}

\def\vL{{\bf L}}
\def\vr{{\bf r}}
\def\rh{\widehat{\bf r}}
\def\sumlz{\sum_{\l=0}^{\infty}}
\def\psih{\widehat{\psi}}

\documentstyle[11pt,epsf,rotate]{article}

\begin{document}


\begin{titlepage}   

\noindent

\begin{center}

\vskip0.9truecm
{\bf

A METHOD FOR EXTRACTING MAXIMUM RESOLUTION POWER SPECTRA 
FROM MICROWAVE SKY MAPS
\footnote{Published in  {\it MNRAS}, {\bf 280}, 299-308 (1996).\\
Available from 
{\it h t t p://www.sns.ias.edu/$\tilde{~}$max/window.html} (faster from the US)\\
and from 
{\it h t t p://www.mpa-garching.mpg.de/$\tilde{~}$max/window.html} (faster from Europe).
}
}

\vskip 0.5truecm
  \auth{Max Tegmark}
  \smallskip

  \addr{Max-Planck-Institut f\"ur Physik, F\"ohringer Ring 6,}
  \addr{D-80805 M\"unchen}

 \addr{email: max@mppmu.mpg.de}

  \smallskip
  \vskip 0.8truecm

{\bf Abstract}
\end{center}
\bigskip
A method for extracting maximal resolution power spectra
from microwave sky maps is presented and applied to 
the 2 year COBE data, yielding a power
spectrum that is consistent with a
standard $n=1$, $\Qrmsps=20\>\mu$K model. 

\noindent
By using weight functions that fall off smoothly near the
galactic cut, it is found that the spectral resolution $\Delta\l$ can be 
more than doubled at $\l=15$ and more than tripled at $\l=20$
compared to simply using galaxy-cut spherical harmonics.
For a future high-resolution experiment with reasonable sky coverage,
the resolution around the CDM Doppler peaks
would be enhanced by a factor of about 100, down 
to $\Delta\l\sim 1$, thus allowing spectral features such as the locations
of the peaks to be determined with great accuracy.
The reason that the improvement is so large is basically that 
functions with a sharp edge at the galaxy cut
exhibit considerable ``ringing" in the Fourier domain,
whereas smooth functions do not. 
The method presented here is applicable to any survey geometry,
chopping strategy and  
exposure pattern whatsoever. 
The so called signal-to-noise eigenfunction 
technique is found to be a special case, corresponding to 
ignoring the width of the window functions.

\end{titlepage}

\section{INTRODUCTION}

There has been a surge of interest in the cosmic microwave 
background radiation (CMB) since the first 
anisotropies of assumed cosmological origin were 
detected by the COBE DMR experiment 
(Smoot {\etal} 1992).
On the experimental front, scores of new experiments have been 
carried out and many more are planned or proposed for 
the near future. 
On the theoretical front, considerable progress 
has been made in understanding 
how the CMB power spectrum $C_\l$ depends on various 
cosmological model parameters -- see Hu \& Sugiyama (1995),
Bond {\etal} 1994 and references therein for recent reviews of
analytical and quantitative aspects of this problem. 
This aim of this paper is to strengthen the link between 
these two fronts, by presenting a method allowing more accurate 
estimation of the power spectrum from experimental data. 

The usual approach to power-spectrum estimation from CMB experiments 
has been to assume that some specific model is true, and 
use a likelihood analysis to find the model parameters that 
best fit the observed data. Since COBE is mainly sensitive to multipoles
$\l\simlt 20$, where many models predict a power spectrum 
$C_\l$ 
that is well fit by a simple power law, most published work 
on the topic has focused on estimating merely two parameters;
the overall normalization and a spectral index $n$, roughly 
the logarithmic slope of the power spectrum.
This approach has since
been generalized to quite a variety of two-parameter 
models, for instance to 
a class of curved $\Omega<1$ cosmologies
(G\'orski {\etal} 1995), where the model parameters are
$\Omega$ and the normalization, and to flat $\Omega<1$ cosmologies
with a cosmological constant (Bunn \& Sugiyama 1995 -- hereafter 
BS95; Stompor {\etal} 1995) where the
model parameters are the cosmological constant $\Lambda$ and 
the normalization. The most elaborate such scheme to date is that of
White \& Bunn (1995, hereafter WB95), using a three-parameter model.
For the purpose of making such parameter fits efficiently, 
a number of sophisticated data analysis techniques have been 
invented, where the data is expanded in some set of 
basis functions that are orthogonal on the celestial 
sphere after the galaxy has been cut out.
It has recently been shown (Tegmark \& Bunn
1995, hereafter referred to as TB95) that the orthogonalized
spherical harmonics (G\'orski 1994, hereafter G94) and the signal-to-noise
eigenmodes (Bond 1994 -- hereafter B95, BS95) 
give results that are both virtually
identical and virtually optimal for parameter fitting,
so that the choice of which method to use is mostly an
issue of computational efficiency.

However, as CMB data continues to improve in quantity and quality, 
it becomes increasingly attractive to go beyond mere
parameter fitting and break the shackles of model-dependence. 
Just as is routinely done with the power spectrum 
$P(k)$ of galaxies, we wish to simply estimate the angular
power spectrum without assuming any model, {\ie}, 
estimate the entire
sequence of multipole moments $C_\l$ from the data one by one, 
and so to say let the data speak for itself.
To estimate say the first 30 multipoles in this fashion thus means
fitting a 30-parameter model to the data. Given the numerical 
difficulties of fitting even three parameters at once, this is 
clearly unfeasible with the above-mentioned methods at the
present time. 

Fortunately, this program can readily be carried through by a 
more direct approach, where the step of maximum-likelihood estimation
is omitted, and the multipoles $C_\l$ are roughly speaking estimated by 
simply expanding the data in some modified versions of the spherical 
harmonics and then squaring the coefficients.
Except for technicalities like shot-noise correction, 
this is analogous to the way that the matter power spectrum 
$P(k)$ is estimated from galaxy redshift surveys: 
the observed density field is expanded in some functions that
resemble plane waves, and then the expansion coefficients are squared.
Pioneering work on this problem (Peebles 1973; 
Hauser \& Peebles 1973) has recently been extended and applied
to the 2 year COBE data (Wright {\etal} 1994b, hereafter W94), 
and the result of such an analysis is shown in 
\fig{PowerFig} (top).
When estimating power spectra,
it is customary (White, Scott \& Silk 1994) 
to place both vertical and horizontal error bars
on the data points, as in \fig{PowerFig}. The former represent
the uncertainty due to noise and cosmic variance, and the latter 
reflect the fact that an estimate of $C_\l$ inadvertently also 
receives contributions from other multipole moments. 
As is well-known, this unavoidable effect is caused by incomplete
sky coverage, which destroys the orthogonality of the
spherical harmonics. 
Another way of phrasing this is that any estimate of
$C_\l$ will be the true power spectrum convolved with a
{\it window function} that has some non-zero width
$\Delta\l$. 
Here and throughout, we will refer to these 
horizontal bars $\Delta\l$ as the {\it resolution} of the 
power spectrum estimate. Poor resolution
(large $\Delta\l$) clearly destroys information by 
smearing out features in the true spectrum.
For typical ground- and balloon-based experiments probing degree
scales, this spectral blurring $\Dl/\l$ tends to be of order unity, 
which makes it difficult to resolve details such as the number of Doppler
peaks. A much better method is that presented in W94, where 
the relative resolution $\Dl/\l$ is brought down to the order 
of $25\%$ by using the COBE sky map. In this paper,
we will see how to reduce the horizontal 
error bars still further, down to their theoretical 
minimum, which for the COBE data is seen to 
be a resolution $\Dl\approx 1$. 
The method is akin in spirit
to that of Tegmark (1995, hereafter T95) 
for power spectrum estimation from 
galaxy surveys, in which the best way to weight the galaxies  
turns out to be given, surprisingly, by the Sch\"odinger equation.  

This paper is organized as follows. 
The problem is formalized in \sec{ProblemSec} and solved in 
\sec{SolutionSec} for the most general case. The 
method is applied
to the special case of the 2 year COBE DMR data in 
\sec{COBEsec}. 
In \sec{QuantumSec}, it is seen that, just as for the above-mentioned
galaxy survey problem, the gist of the method can be
understood from a simple analogy with quantum mechanics.
The method is compared to other techniques in \sec{DiscussionSec}, and 
its implications for future CMB experiments are discussed.

\section{THE PROBLEM}
\label{ProblemSec}

Although the following three sections are rather technical in nature,
focusing on the application to real data, 
we will see in \sec{QuantumSec} that the essence of both the problem and
its solution have quite a simple and intuitive interpretation.

Since the computations that follow are all linear algebra,
vector and matrix notation will be used as extensively as possible.
If the introduction of some notation is found too
cursory, the reader is referred to TB95 for more details. 
Although the true CMB fluctuations are described by some
smooth temperature distribution ${\Delta T\over T}$ 
over the celestial sphere, 
a real-world CMB sky map is always discretized into some 
finite number of pixels, say $N$ of them.
Let us write the CMB sky map as the $N$-dimensional 
vector ${\vx}$, where 
\beq{xDefEq}
x_i = {\Delta T\over T}(\rh_i),
\eeq
and $\rh_i$ is a unit vector 
in the direction of the $i^{th}$ pixel.
$\N=6144$ for the all-sky COBE DMR map.

For reasons that will become clear later, we define
a new vector 
$\vz\equiv A\vx$, where $A$ is some completely arbitrary 
$N'\times N$ matrix. Later on, we will see that considerable 
savings in computer time 
can often be made by choosing $A$ wisely.
Using equation (2) from TB95, we obtain
\beq{zEq}
\vz = A(Y\va+\vn)
\eeq
by choosing the matrix $A$ to have all its row vectors
orthogonal to the monopole and the dipole, thus 
eliminating the ``nuisance contribution" from these multipoles.
The $N\times\infty$-dimensional spherical harmonic matrix $Y$ 
is defined as 
$Y_{i\lambda} \equiv Y_{\l m}(\rh_i)$,
where we have combined $\l$ and $m$ into the single index 
$\lambda\equiv \l^2+\l+m+1 = 1, 2, 3, ...$.
(Throughout this paper, we use real-valued spherical harmonics, which
are obtained from the standard spherical harmonics by replacing $e^{im\phi}$
by $\sqrt 2\sin m\phi$, $1$, $\sqrt 2\cos m\phi$ for $m<0$, 
$m=0$, $m>0$
respectively.)
Making the standard assumption that 
the CMB is Gaussian on the scales probed, $\va$ is an
infinite-dimensional Gaussian random vector
with zero mean and with the diagonal covariance matrix
\beq{Cldef}
\expec{a_{\lambda}a_{\lambda'}} = \delta_{\lambda\lambda'} B_\l^2 C_\l,
\eeq
where $C_\l$ is the angular power spectrum
and where $B_\l^2$ is the experimental beam function. 
The $\N$-dimensional
noise
vector $\vn$ is assumed to be Gaussian with 
$\expec{\vn} = 0$ and  
$\expec{\n_i\n_j} = \sigma_i\sigma_j \delta_{ij}$
(for the COBE case, see Lineweaver {\etal} 1994), 
where $\sigma_i$ is the rms noise of pixel $i$.

We wish to estimate the power spectrum $C_\l$ from $\vz$. 
As $C_\l$ tends to decrease rapidly with $\l$, we will 
find it convenient to define
\beq{DdefEq}
D_\l \equiv B_\l^2 C_\l/\mu_\l
\eeq
for some weights $\mu_\l$ that make the coefficients 
$D_{\l}$ of similar magnitude, as was done in WB95.
We will leave $\mu_{\l}$ arbitrary for now, and 
estimate $\D_{\l}$.
Everything below refers to some definite multipole $\l^*$, say $\l^*=7$, 
and is to be repeated separately for all other $\l^*$-values of interest. 
The most general estimate of $\D_{\l^*}$ that is quadratic in
the data can clearly
be written as 
\beq{DestDefEq}
\Dest\equiv\vz^t\El\vz,
\eeq
where $\El$ is some arbitrary real symmetric 
$N'\times N'$ matrix. 
Let us expand it
in an orthogonal set of eigenvectors $\{\vek\}$ as
\beq{EexpansionEq}
\El = \sumk\weight_k\vek\vek^t,
\eeq
where $\weight_k \neq 0$ (we simply omit vanishing 
eigenvalues from the sum, so that
$N''$ is the rank of the matrix).
Since $\D$ is by definition positive, we clearly want 
$\El$ to be positive semidefinite, {\ie},
$\weight_k>0$.
Substituting 
equations\eqnum{zEq},\eqnum{DdefEq} and\eqnum{EexpansionEq}
into\eqnum{DestDefEq}, we obtain
\beqa{GeneralWindowEq}
\expec{\Dest} &=& 
\sumk\weight_k\left[
\tr Y^t A^t\vek\vek^t AY\expec{\va\va^t}
+ \tr A^t\vek\vek^t A \expec{\vn\vn^t}\right]\\
&=& \sumlp\v_{\l}^2 D_{\l} + \bias,
\eeqa
where we have defined the {\it window function}
\beq{WindowEq}
\v_{\l}^2 \equiv 
\mu_\l\sumk\weight_k\summp
(V^t\vek)_{\lambda}^2
\eeq
and the {\it noise bias}
\beq{biasEq}
\bias\equiv\sumk\weight_k\sum_{i=1}^N
(A^t\vek)_i^2\sigma_i^2,
\eeq
and $V\equiv AY$. 
This merely reflects well-known fact that 
the expectation value of
any quadratic combination of pixels
is just the power spectrum convolved with some 
window function, plus a positive noise bias.
We define the {\it noise-corrected} estimate by 
\beq{NedWrightEq}
\Dest^{corr} \equiv \Dest - B.
\eeq
We want the window function to represent a weighted average of the
$D_{\l}$-coefficients, which requires it to be normalized so that
\beq{WinNormEq}
\sumlp \v_{\l}^2 = 1.
\eeq
Defining 
\beq{NormEq}
\expec{f(\l)}\equiv \sumlp \v_{\l}^2 f(\l) 
\eeq
for an arbitrary function $f$, 
we can write this normalization condition as simply $\expec{1}=1$.
Ideally, we would want the window function to be 
$\v_{\l}^2 = \delta_{\l\l^*}$, but this is of course
impossible to achieve if 
the sky coverage is incomplete. Given this limitation, we simply want 
the window function to be {\it as narrow as possible}, 
centered around $\l^*$. We may for instance choose to minimize 
its variance (second moment) about $\l^*$, $\expec{(\l-\l^*)^2}$,
or more generally the quantity $\expec{|\l-\l^*|^{\nu}}$ 
for some positive exponent 
$\nu$. Let us allow for complete freedom of preferences by minimizing
$\expec{\u_{\l}}$ for a completely arbitrary {\it penalty function}
$\u_{\l}$. We thus arrive at the following optimization problem:

\noindent
Find the matrix $\El$ that minimizes $\expec{\u_{\l}}$ 
subject to the constraint that $\expec{1}=1$.

\section{THE SOLUTION}
\label{SolutionSec}

We will now solve this constrained minimization problem. 
It is easy to see that an optimal solution can always be found
where the matrix $\El$ has rank 1, {\ie}, where it is of the 
simple form $\El =\ve\ve^t$ for some vector $\ve$. 
Below we will find a set of $N'$ different rank 1 matrices
$\vek\vek^t$ which have the 
following properties:
\begin{itemize}

\item They are independent in the sense that
$\expec{(\vek\cdot\vz)(\vekp\cdot\vz)} \propto \delta_{kk'}$.

\item
$k=1$ corresponds to an optimal solution, $k=2$ to the second best solution,
{\etc}

\end{itemize}

\noindent
If the only objective were to minimize the horizontal error bars in
\fig{PowerFig}, the optimal choice would thus be 
$\El = \veone\veone^t$.
However, we clearly also want to keep the vertical error bars as small
as we can. Since cosmic variance drops as 
we average many different estimates, it is in general better to 
choose $\El$ to be some weighted average of the above-mentioned
rank one matrices, as given by \eq{EexpansionEq} when normalizing
the weights as 
$\sumk\weight_k = 1$.
We clearly face a trade-off between vertical and horizontal 
error bars: as we increase $N''$, the vertical error bar decreases,
but since we are including increasingly wide window functions,
the horizontal error bar increases.
For reasonable sky coverage except for a galactic cut, the choice of $N''$ 
turns out to be quite an easy one:
the best $(2\l^*+1)$ rank one matrices clearly stand out in front of the rest
of the pack, and all give quite similar error bars. Once we have fixed
$N''$, changing the weights $\weight_k$ leaves the horizontal
error bars fairly unaffected, and we can choose $\weight_k$ 
so as to minimize the vertical error bars.
In the limit of complete
sky coverage, these $(2\l^*+1)$ best vectors $\vek$ approach
the $(2\l^*+1)$ spherical harmonics in question, just as we would expect.

Now let us find the vectors $\vek$.
Writing $\El = \vek\vek^t$ (no summation implied), 
we solve the constrained minimization 
problem for $\vek$ using the 
standard method of Lagrange multipliers.
Defining
\beq{LdefEq}
\Ell \equiv \expec{\u_{\l}} + \scale B - \lambda(\expec{1}-1)
\eeq
and requiring the derivatives with respect to the components of 
$\vek$ to vanish, we obtain
\beq{EigenEq}
(Q - \lambda R)\vek = 0,
\eeq
where we have defined the $N'\times N'$ matrices
\beqa{QdefEq}
Q_{ij}&=&\sum_{\lambda}V_{i\lambda}V_{j\lambda}\mu_{\l}\u_{\l}
+ \scale\sum_{k=1}^N A_{ik} A_{jk}\sigma_k^2,\\
\label{RdefEq}
R_{ij}&=&\sum_{\lambda}V_{i\lambda}V_{j\lambda}\mu_{\l}.
\eeqa
Here we have added the term $\scale B$ to the target function
to make the noise bias $B$ (and thus the vertical 
error bars) small. The choice of $\scale$ is discussed below.
Since both $Q$ and $R$ are symmetric, 
this generalized eigenvalue problem has $N'$ real 
solutions, which we normalize so that 
$\expec{1}=1$ and
sort by their eigenvalues
(the smallest eigenvalue corresponds to $k=1$, the best solution, {\etc}).
Most standard eigenvalue packages (such as the public-domain 
package EISPACK, or that included in NAG) 
provide a specialized routine for  
precisely this problem: the generalized eigenvalue problem 
where $Q$ and $R$ are real and symmetric, and $R$ is positive
definite.

Once we have solved this and chosen $N''$, the variance of our 
estimate is
\beq{VarianceEq}
V(\Dest) = 2\sum_k\sum_{k'} \weight_k\weight_{k'}
M_{kk'}^2,
\eeq
where the matrix 
$M_{kk'}\equiv\expec{(\vek\cdot z)(\vekp\cdot z)}$ is given by
\beq{Meq}
M_{kk'} = 
\sum_{\lambda} 
(V^t \vek)_{\lambda} (V^t \vekp)_{\lambda}
\mu_\l D_\l + 
\sum_{i=1}^N
(A^t \vek)_i (A^t \vekp)_i
\sigma^2_i.
\eeq
When $M$ is almost diagonal, 
the variance is approximately minimized if we choose 
the weights
$\weight_k \propto M_{kk}^{-2}$.
It is easy to show that the first term will be strictly diagonal.
The second term will typically be close to diagonal, with
a slight off-diagonality caused by heteroscedastic noise.

\subsection{How to chose the parameters}

The above treatment was very general, leaving the choice of 
the matrix $A$, the weights $\mu$, the penalty function
$\u_\l$ and the constant $\scale$ to be chosen according to 
the preferences of the user.
We now discuss some aspects of the
choice of these free parameters.

{\bf The matrix $A$.}
$A$ is conveniently constructed in two steps: 
\begin{enumerate}

\item 
One makes some natural choice such as the spherical 
harmonic matrix $Y^t$, the noise-weighted and
orthogonalized ditto introduced by G94,
the time-saving matrix described
in the next section, or the identity matrix if one prefers 
to simply use the pixel values 
for a brute-force approach \`a la TB95, obtaining the best possible
result at a high computational cost.

\item
One makes all the row vectors orthogonal to the monopole and the dipole,
as described in {\eg} TB95.

\end{enumerate}
The matrix $A$ also allows us to directly incorporate data sets
other than sky maps. For example, for a 
double-chop configuration such as that of the Tenerife experiment,
the observed data vector $\vz$ is a certain linear combination of 
the actual sky pixels $\vx$ (which are never observed directly), and since
the coefficients in these linear combinations (the elements 
of $A$) are known, the matrix $A$ incorporates all we need to know about
the data acquisition process.

{\bf The weights $\mu_\l$.}
This is basically the choice of what to label the vertical axis with
in \fig{PowerFig}. 
If we probe a linear function with a symmetric 
and correctly centered window function, the estimates will be unbiased. 
However, if we are probing the power spectrum $C_\l$, which 
is generally believed to have convex shape 
(the second derivative of say $1/\l(\l+1)$ is positive), then
our estimates tend to be biased high, since 
the upward bias from coupling to lower multipoles (the so called 
``red leak")
is stronger than the downward bias from coupling to higher multipoles
(the ``blue leak").
For instance, when attempting to estimate $C_{20}$,
even a very slight sensitivity to the quadrupole $C_2$ is likely 
to dwarf the signal one is trying to measure, as pointed out in
{\eg} W94 and illustrated in \fig{HistFig}.
A simple strategy for reducing such problems
is to choose $\mu_\l$ proportional to the expected power spectrum 
multiplied by the experimental window function, \ie, 
\beq{muChoiceEq}
\mu_\l = {B_\l^2\over \l(\l+1)},
\eeq
so that 
the coefficients $D_\l$ will all be of the same order of magnitude.
This is quite a natural choice also because power spectra are 
conventionally plotted with the quantity 
$\l(\l+1)C_\l/2\pi\propto D_\l$ 
on the vertical axis. 
Thus the horizontal 
error bars will refer to exactly the 
quantities that we plot, {\ie}, $D_\l$,
which is of course the most honest way to present 
the results.

{\bf The penalty function $\u$.}
The simple choice $\u_{\l} = (\l-\l^*)^2$ is quite 
adequate if $\mu_\l$ is a mere constant.
This gives $p_\l\mu_\l\sim\l^2$ as $\l\to\infty$, 
and corresponds to 
the quadratic penalty function $\u_{\vk} = k^2$ as $\l\to\infty$
in the galaxy survey problem treated in T95. 
In T95, it was seen that this $k^2$ penalty in 
Fourier space corresponded to a Laplace operator
$\nabla^2$ in real space, and that this 
guaranteed that the resulting solutions were 
continuous and well-behaved.
A $k^4$ term would make the solutions even smoother, guaranteeing
that all derivatives would be continuous as well, {\etc}
If $p_\l\mu_\l$ asymptotically grows slower than $\l^2$, 
however, high frequencies in the weight function 
are not sufficiently suppressed, and there is a risk that the 
solutions will be quite irregular and unnatural. 
Since we advocated choosing 
$\mu_l = B_\l^2/\l(\l+1)$ above, 
we thus need to make $\u_\l\to\infty$
at least as fast as $\l^4/B_\l^2$ as $\l\to\infty$. 
\noindent
In addition, if one is particularly worried about say
non-cosmic contamination from a particular multipole like the 
quadrupole,
one can increase the value of $p_2$, thereby reducing the
value of the window function at $\l=2$. 
There are of course no free lunches: the price of 
reducing $v_2^2$ in this fashion is that other unwanted
window function entries increase. 

{\bf The constant $\bf\scale$.}
Since there is a trade-off between vertical and horizontal
error bars, we minimize a combination of them.
The larger we make $\scale$, the greater the emphasis on the 
vertical ones. A larger $\scale$ tends to increase the weights 
given to the least noisy pixels.
As long as the pixel noise levels 
are of the same order of magnitude all across the map, 
the results are quite insensitive to 
the choice of $\scale$, and one might as well set 
$\scale=0$.

It should be stressed that these free parameters reflect 
a strength rather than
a weakness of the method. The goal is to find 
optimal window functions, and the choices of $\mu$, $\u$ 
and $\scale$
simply reflect what we mean by ``good". Different choices give slightly
different error bars on the resulting plots, but even if one criterion
for ``good" is used in the optimization process and a different
one is used when judging the results,
the optimization tends to improve the situation greatly over 
simply using spherical harmonics. In \sec{QuantumSec}, we will see why.

\subsection{Analyzing very large data sets}

When applying this method to COBE, where the matrices are never
larger than $4038\times 4038$, the solution of the
resulting eigenvalue problem is numerically 
straightforward. 
Future high resolution 
CMB experiments, however, are likely to produce data sets where the
pixels are counted in millions rather than thousands.
Although it is obviously not feasible to diagonalize matrices 
of this size, the method presented above can nonetheless be readily 
applied to such data sets, as follows:
\begin{itemize}

\item
For small $\l$, almost no information is lost if the number of
pixels is reduced by smoothing prior to the analysis.

\item
For large $\l$, it is convenient to split the sky into 
many smaller patches, analyze them separately, and then combine 
the results. This of course lowers the spectral 
resolution somewhat, but as is discussed in \sec{QuantumSec}, 
$\Dl/\l$ can still be kept extremely small, of the order of a few
percent.

\end{itemize}

\section{APPLICATION TO COBE}
\label{COBEsec}

We now apply the method to the two year COBE DMR data, combining 
the 53 and 90 GHz channels and removing all pixels less than 
$20^{\circ}$ away from the galactic plane. This leaves $N=4038$ pixels.

\subsection{Parameter choices}

Just as in WB95, we choose 
$\mu_\l = B_\l^2/\l(\l+1)$, where $B_\l^2$ is the
COBE beam window function published by Wright {\etal} (1994a). 
In line with our discussion above, we estimate $D_{\l^*}$
by chosing the penalty function 
\beq{PenaltyChoiceEq}
\u_\l = \cases{
(\l-\l^*)^2/\mu_{\l^*+2}   	&if $\l\le\l^*+2$,\crr
(\l-\l^*)^2/\mu_{\l}		&if $\l\ge\l^*+2$.\crr
}
\eeq
This behaves as a
simple quadratic penalty function
$\u_{\l} \propto (\l-\l^*)^2$ near $\l^*$, whose symmetry
ensures that
the ``red leak" will be similar to the ``blue leak" and
thus that $\expec{\l}\approx\l^*$, but
also has the above-mentioned desired property 
that $\u_{\l}\mu_\l\propto\l^2$
as $\l\to\infty$.
As the COBE signal-to-noise is quite good and the pixel noise 
does not vary radically across the sky, we make the simple choice $\scale=0$.
As the noise bias $B$ is of order 
$B_*\equiv (70\mK)^2 \l(\l+1)/N B_\l^2$ and the 
optimal horizontal error bars $\Delta\l$ are of order unity, a 
natural alternative choice would be say
$\scale = 1/5 B_*$. This choice would mean roughly that 
we value a $1\%$ reduction in the horizontal error bar as much  
as a $5\%$ reduction in the vertical error bar.
We choose $A$ to be the first $961$ rows of $Y^t$, which corresponds to 
the spherical harmonic components up to $\l=30$. 
We place the cutoff at $\l=30$ simply because the 
COBE-beam effectively washes out the signal from
$\l\gg20$, leaving mostly noise at these high frequencies.
Indeed, by 
comparing G94 and TB95, 
one concludes that 
$\vz$ contains almost all the cosmological
information in $\vx$ with this high frequency cutoff.
We then throw out the first four rows of $A$, corresponding
to the monopole and dipole, and make the remaining rows orthogonal to them.

\subsection{A trick for saving time}

Because the galactic cut preserves reflection symmetry about the galactic
plane, even and odd multipoles remain orthogonal to one another 
(to very good accuracy, the only slight correction arising from  
pixelization effects). When estimating the $\D_{\l}$ corresponding
to an even $\l$, we thus throw out all rows corresponding to odd $\l$, 
and vice versa. 

Although diagonalizing $500\times 500$ matrices ($N'$ being of order
$961/2$) is numerically
straightforward, the solution of the eigenvalue problem 
given by \eq{EigenEq} can be simplified further.
Since the galactic cut preserves azimuthal symmetry, spherical 
harmonics with different $m$-values are also orthogonal to each other,
so by simply regrouping the rows of $A$ by $m$-values,
the matrices $Q$ and $R$ become block-diagonal.
In other words, we can do the following:

\begin{enumerate}

\item Pick some $m$-value and choose $A$ to be the rows 
of $Y^t$ that are not 
orthogonal to the row corresponding to $Y_{\l,m}$.
For say $m=0$ and $\l^*$ even, $A$ will consist of the 
$N'=14$ rows corresponding to 
$Y_{2,m}$, $Y_{4,m}$, ... , $Y_{30,m}$.

\item
Compute $Q$ and $R$, truncating the 
sums \eqnum{QdefEq} and \eqnum{RdefEq} at say $\l=60$,
since the COBE beam $B_{\l}$ exponentially suppresses 
multipoles $\l\gg 15$. 

\item
Find the smallest eigenvalue of \eq{EigenEq},
normalize the corresponding eigenvector so that 
$\expec{1}=1$ and denote it $\ve_m$.

\item
Repeat the corresponding procedure for 
the other $m$-values, finally obtaining the $N''=2\l^*+1$ 
vectors $\{\ve_m\}$.

\item Compute the matrix $M$ (as with all other methods,
error bar estimation of course requires 
assuming some reasonable power spectrum --- here 
the spectrum $n=1$, $\Qrmsps=20\mK$ is used) 
and the optimal weights $\alpha_m$.

\item Compute $\Dest^{corr}$ and the vertical and horizontal
error bars.

\item Repeat everything for all other $\l^*$-values of interest.

\end{enumerate}

\noindent
This is what has been done to produce \fig{PowerFig} (bottom).

\subsection{Results}

\begin{figure}[phbt]
\vskip-3cm
\centerline{\epsfxsize=18cm\epsfbox{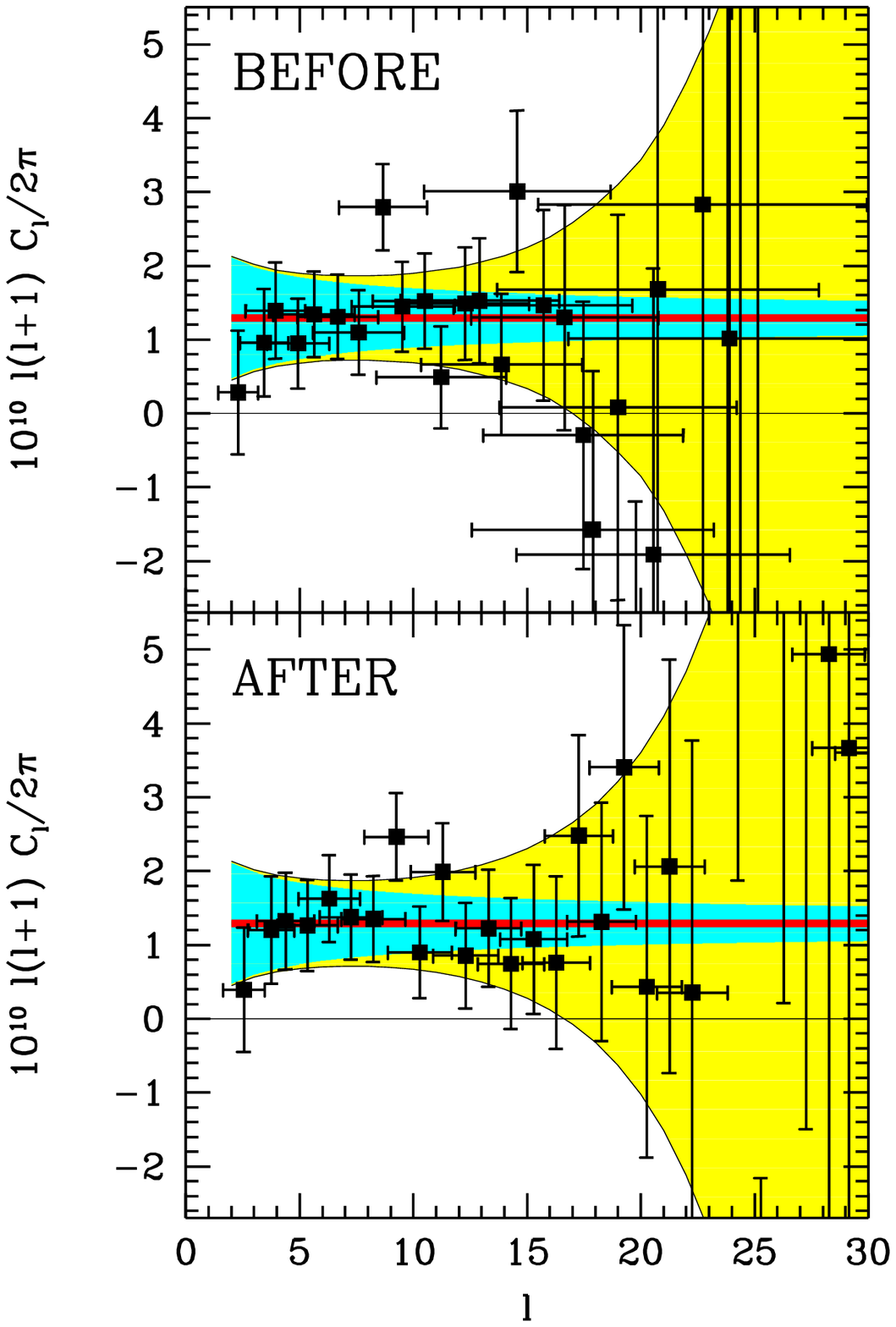}}
\vskip-1cm
\caption{
The COBE power spectrum before and after optimization.
}
The observed multipoles $D_\l = \l(\l+1)C_\l$ are plotted with
$1-\sigma$ error bars. The vertical error bars include both
pixel noise and cosmic variance, and the horizontal error bars 
show the width of the window functions used. 
If the true power spectrum is given by 
$n=1$ and $\Qrmsps=20\mK$ (the heavy horizontal line), 
then the shaded region gives the $1-\sigma$ error bars and
the dark-shaded region shows the contribution from
cosmic variance. 
The multipoles have been estimated by expansion in 
galaxy-cut spherical harmonics (top) and with the optimal 
method (bottom).
\label{PowerFig}
\end{figure}

As expected, \fig{PowerFig} (bottom) shows the vertical error bars 
being dominated by cosmic variance 
for low $\l$ and by noise for large $\l$,
especially for $\l\simgt 15$, where 
beam smearing takes its heavy toll.
As to the horizontal error bars, the bars that specify
the spectral resolution, we see that they
exhibit hardly any $\l$-dependence --- we will return
to this fact in the 
next section, and see that it has a simple intuitive 
explanation.
Having said this about the error bars, we now turn 
to the location of the data points.
If the true power spectrum $C_\l$ were a simple $n=1$ 
Harrison-Zel'dovich spectrum with normalization 
$\Qrmsps = 20\mK$, then we would expect about 
$68\%$ of the data points in \fig{PowerFig} to 
lie within the shaded region, which is the 
$1-\sigma$ error region for this model, and this is clearly the
case, in agreement with previous analyses such as W94
and de Oliveira-Costa \& Smoot (1995).
Comparing our results to those obtained with the 
truncated spherical harmonic method (top),
we see that the 
individual multipole estimates agree quite well for low
$\l$, where the spectral resolution $\Dl$ of the two methods
is comparable, but start to differ as $\l$ increases, reflecting
the fact that the data points in the upper figure are 
probing the average power in a much wider band of $\l$-values.

For verification, 
1000 Monte-Carlo simulations were 
performed, where fake COBE-skies with noise and galaxy cut 
were generated with $n=1$, 
$\Qrmsps = 20\mK$ and then piped though the analysis software.
The result was in perfect agreement with the theoretically
predicted error bars, with zero bias, 
about $68\%$ of the estimated multipoles falling within the shaded
region in \fig{PowerFig}, {\etc}
The COBE analysis was also repeated for a number of different choices of
the parameters $\scale$, $\mu$ and $\u$, and the results were found to 
remain virtually
unaffected. In other words, the results obtained with the
method we have presented are quite robust.

\section{THE QUANTUM ANALOGY}
\label{QuantumSec}

Whereas sections\secnum{ProblemSec},\secnum{SolutionSec}
and\secnum{COBEsec} of this paper were rather technical, 
this section shows
that all qualitative features of the 
method can be understood from a simple analogy with quantum mechanics.
Although all the elements of realism 
included above
are important when applying the method
to real data, one can in fact 
understand the qualitative features
by ignoring most of these complications.

First of all, let us ignore the fact that 
a real CMB sky map is pixelized. We let 
the function $x(\rh)$ denote the temperature fluctuation $\Delta T/T$ in the 
direction of the unit vector $\rh$. 
In the discrete real-world case, 
each pixel has some r.m.s. noise variance $\sigma^2$ and 
effectively covers some solid angle $\Omega$ of the sky. 
For our continuous analogue, let us define the noise variance function $V(\rh)$
by $V\equiv\sigma^2\Omega$, where $\sigma$ and $\Omega$ refer to a pixel in the
direction $\rh$. 
If we neglect contamination problems, it is easy to see that $V(\rh)$ is simply 
proportional to the inverse of the
time spent observing the direction $\rh$ (the small-scale details of the pointing 
of the antenna are irrelevant, as the beam smearing will ensure
that the function $V$ is smooth on angular scales below the beam width).
Secondly, let us ignore the nuisance 
terms from monopole, dipole, {\etc}, so that we
can omit $A$ from \eq{zEq}.
Finally, we choose $\mu_\l = 1$.

Since the power $D_{\l^*}$ has units of $\mK^2$, we clearly want to estimate
it by some quantity $\Dt$ that is quadratic in the data. 
The simplest estimator of this type is 
\beq{EstDefEq}
\Dt \equiv \left|\int \psi(\rh)x(\rh)d\Omega\right|^2,
\eeq
where $\psi$ is some function on the sphere,
and above we showed that the most general estimate is 
just a weighted average of such estimates. 
A straightforward calculation shows that 
\beq{CestExpecEq}
\expec{\Dt} = 
\sumlz\summ |\psih_{lm}|^2 D_\l + \int|\psi(\rh)|^2 V(\rh) d\Omega,
\eeq
where $\psih_{lm}$ denotes the spherical Fourier transform of $\psi$, {\ie}, 
the coefficients in an expansion of $\psi$ in spherical
harmonics. 
In other words, we see that the expectation value of our estimator
is the sum of two terms of quite different character.
The first, the contribution from cosmology, 
is the power spectrum convolved with 
a window function 
$v_\l^2\equiv\sum_m |\psih_{lm}|^2$.
The second, the contribution from noise, is just an average value 
of $V$, the weights being $|\psi(\rh)|^2$.
This is very similar to the result when estimating the power
spectrum from a galaxy survey (T95).
Just as in that paper, we will find it
convenient to use the
standard Dirac quantum mechanics notation with kets,
bras and linear operators. 
This allows us to write $\psih_{\l m} = \expec{\l m|\psi}$.
A window function should always integrate to unity, 
so the correct normalization 
for $\psi$ is just $\expec{\psi|\psi}=1$.
Defining the operator 
\beq{LopDefEq}
L\equiv\sumlz\summ \l|\l m\rangle\langle\l m|,
\eeq
\eq{CestExpecEq} becomes simply
\beq{CestExpecEq2}
\expec{\Dt} = \expec{\psi|D_L + V(\rh)|\psi}.
\eeq
Note that $L$ is a scalar operator satisfying 
$L|\l m\rangle = \l |\l m\rangle$, and is related to the 
(vector) angular momentum operator 
$\vL = -i\vr\times\nabla$
through $\vL^2 = L(L+1)$. 

Now what is the best choice of $\psi$?
Equation\eqnum{EstDefEq} tells us that $\Dt$ is the square modulus
of a random variable whose real and imaginary parts are both Gaussian. 
Thus if $\psi$ is real, the standard deviation of $\Dt$
(the vertical error bar) is simply $\sqrt{2}$ times its expectation value. 
(With random phases, the real and imaginary parts contribute equally, 
and this decreases the variance by a factor of $\sqrt{2}$.)
This means that we minimize the vertical error bars by minimizing 
$\expec{\Dt}$. But assuming that our window function is narrow
enough that we are measuring mostly what we want to measure, 
$D_{\l^*}$, 
the first term in \eq{CestExpecEq2} satisfies
$\expec{\psi|D_L|\psi} \approx D_{\l^*}$, independent
of $\psi$, so we minimize the vertical error bars by simply minimizing the 
second term, $\expec{\psi|V(\rh)|\psi}$.
As a measure of the horizontal error bars, we will use
$\Dl$, the r.m.s. deviation
of the window function from $\l^*$.
With our quantum notation, we have simply 
$\Dl^2 = \expec{\psi|(L-\l^*)^2|\psi}$.
As was discussed above, it is impossible to minimize 
both error bars at
the same time, since there is a trade-off between them. 
It would be like asking for the best and cheapest car.
Instead, the best we can do is minimize some linear combination 
$E \equiv \expec{H}$, where we have defined
\beq{HamiltonianEq}
H \equiv (L-\l^*)^2 + \scale V(\rh),
\eeq
and the parameter $\scale$ specifies how concerned we are about the
vertical error bar relative to the horizontal one. 
Continuing our quantum analogy, we see that we want to find
the $\psi$ that minimizes the total ``energy", 
where the ``kinetic energy" $(L-\l^*)^2$ corresponds
to the horizontal error bar and the ``potential energy" 
$\scale V(\rh)$ corresponds
to the vertical error bar.
If we for the sake of illustration 
set $\l^*=-1/2$, we simply want to minimize
$\expec{\psi|\vL^2 + \scale V(\rh)|\psi}$, given the constraint
$\expec{\psi|\psi}=1$.
Introducing a Lagrange multiplier $E$, we arrive at the Schr\"odinger equation
\beq{ErwinEq}
[\vL^2 + \scale V(\rh)]|\psi\rangle = E|\psi\rangle.
\eeq
In other words, we want to find the ground state wavefunction for a particle
confined to a sphere with some potential. 
Numerically solving for various integer values of $\l^*$
gives functions with similar behavior, modulated by a 
wiggling similar to that of the corresponding spherical harmonics.

\begin{figure}[phbt]
\vskip-2.5cm
\centerline{\epsfxsize=16cm\epsfbox{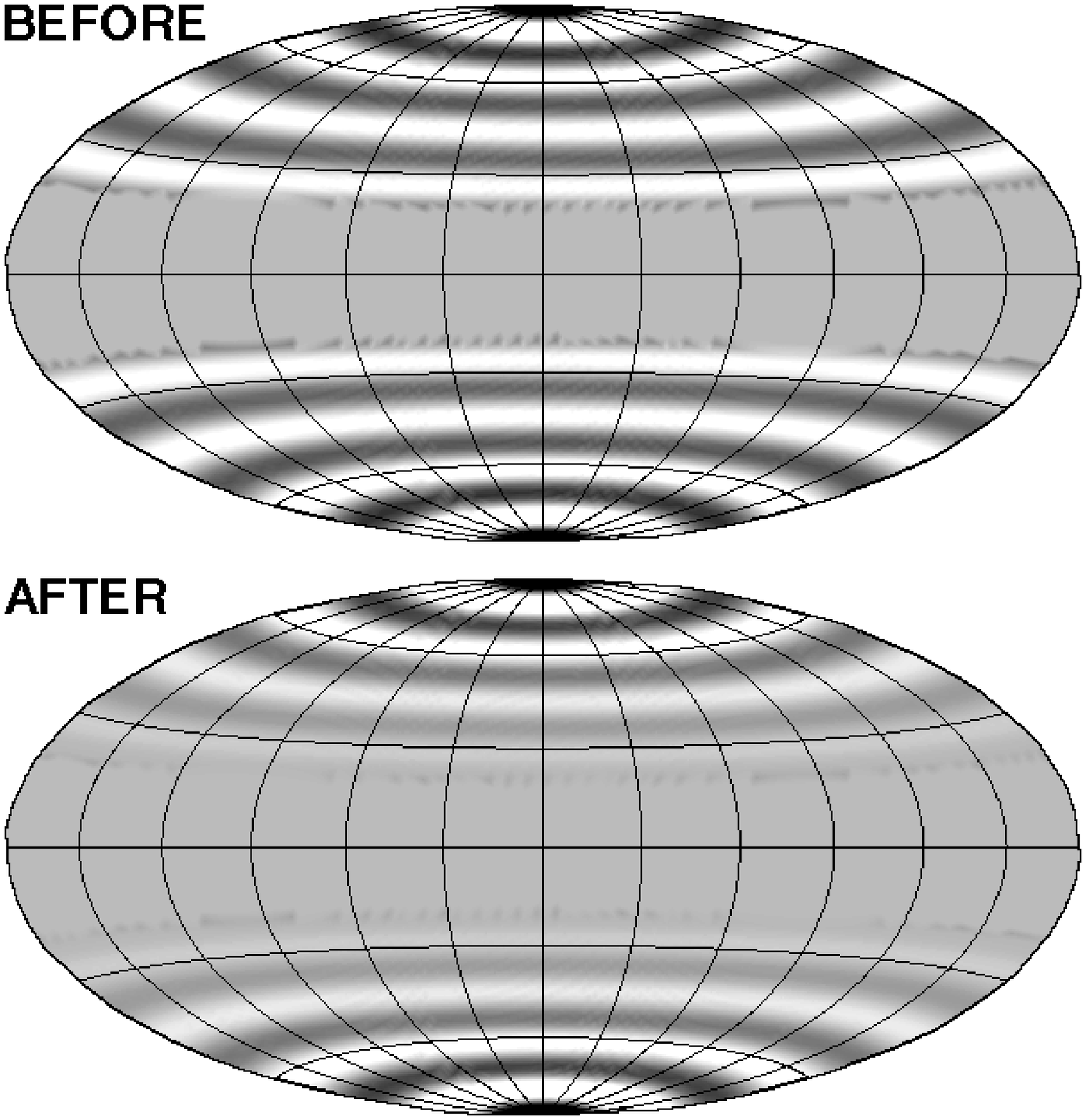}}
\caption{
Weight functions before and after optimization.
}
The weight function ${\bf e}$ used to estimate
the multipole $\l=20$, $m=0$ is plotted in 
Hammer-Aitoff projection in galactic coordinates.
Black pixels receive large positive weight,
white pixels receive large negative weight, and 
the grey shade of the galactic
cut corresponds to zero weight.
The functions shown are the relevant 
galaxy-cut spherical harmonic (top) and 
the eigenfunction ${\bf e}$ with the smallest 
eigenvalue (bottom).
\label{MapFig}
\end{figure}

From our knowledge of quantum mechanics, 
we can immediately draw a 
number of conclusions about the solutions, all which turn out to agree
well with the exact numerical results.
\begin{itemize}

\item $|\psi|^2$ will be small in regions where the noise variance is large,
so regions that received little observation time will receive low
weights in the analysis.

\ns
\ns
\item Except for the case of complete sky coverage, we will have $\Dl>0$. 

\ns
\ns
\item If incomplete sky coverage confines $\psi$ to a region of the sky whose
angular diameter in the narrowest direction is of order $\Delta\theta$, then 
the uncertainty principle tells us that the minimum $\Dl$ must be at least
of order $1/\Delta\theta$. 

\ns
\ns
\item This limit on the spectral resolution is independent of $\l$
(that this remains approximately true in the real-world case as 
well is illustrated in \Fig{ComparisonFig}). 

\ns
\ns
For a sky map of COBE type, where $\Delta\theta$ is of order a radian
given a $20^{\circ}$ galactic cut,
the uncertainty principle thus gives $\Delta\l\simgt 1$. This
agrees well with the horizontal error bars actually attained in 
Figure 1 (bottom).

\ns
\ns
\item If the sky-coverage is incomplete, $V$ is infinite outside of the region
covered, and we recover the quantum-mechanical particle-in-a-box problem.
From this we know that $\psi$ will always go to zero smoothly as it approaches the
survey boundary. This is illustrated in 
\fig{MapFig} (bottom) by a numerical example.

\ns
\ns
\item This smoothness of $\psi$ is really the gist of the method,
as it radically reduces ``ringing" in Fourier space,
``kinetic energy", without increasing 
the ``potential energy" $\expec{\psi|V(\rh)|\psi}$ much at all.
This is seen by comparing figures\fignum{MapFig} and \fignum{HistFig}:
whereas the upper and lower weight functions look 
quite similar in real space (\fig{MapFig}), they differ
dramatically in the Fourier (multipole) domain (\fig{HistFig}).

\ns
\ns
\item If the noise level $\V$ is constant wherever we have data, 
then the potential 
energy term will reduce to $\scale\expec{\psi|V|\psi} = 
\scale V\expec{\psi|\psi} = \scale V$, {\ie}, become independent
of $\psi$. This means that the solution will be independent of $\scale$. 
For any evenly sampled data set, the choice of $\scale$ is thus
irrelevant, so we might as well chose $\scale=0$ for simplicity,
as we did in our COBE analysis. 

\ns
\ns
\item If the survey volume consists of several disconnected
parts, then $\Dl$ is limited by the $\Delta\theta$ of the largest part. 
For the galaxy-cut COBE case, for instance, using only the northern 
half of the sky gives almost the same 
$\Dl$ as using both the northern and southern skies combined.
(However, including both of course helps 
reduce the {\it vertical} error bars.)

\end{itemize}

\section{DISCUSSION}
\label{DiscussionSec}

A method for extracting maximal resolution power spectra from 
CMB sky maps has been presented and applied to the 2 year COBE DMR data.

\subsection{COBE results}

The power spectrum extracted from the 2 year COBE data in \fig{PowerFig}
(bottom) is seen to be consistent with the standard 
$n=1$, $\Qrmsps=20$ model.
This model is close to the best-fit models found in
the various two-parameter Bayesian likelihood analyses published 
(\eg, G94, B95, BS95, TB95, WB95), which we can interpret
as the best-fit straight line through the data points 
in \fig{PowerFig} being close to the 
horizontal heavy line.
As has frequently been pointed out (see {\eg} WB95),
Bayesean methods by their very nature can only make 
{\it relative} statements of merit about different models,
and never address the question of whether the best-fit model
itself is in fact inconsistent with the data.
As an absurd example, the best fit straight line to a 
parabola on the interval $[-1,1]$ is horizontal, even though
this is a terrible fit to the data. It is thus quite 
reassuring that the power spectrum in \fig{PowerFig}
(bottom) not only has the right average normalization and
slope, but that each and every one of the multipoles 
appear to be consistent with this standard best fit model.

\subsection{How the method differs from other techniques}

\begin{figure}[phbt]
\centerline{\rotate[r]{\vbox{\epsfysize=14cm\epsfysize=14cm\epsfbox{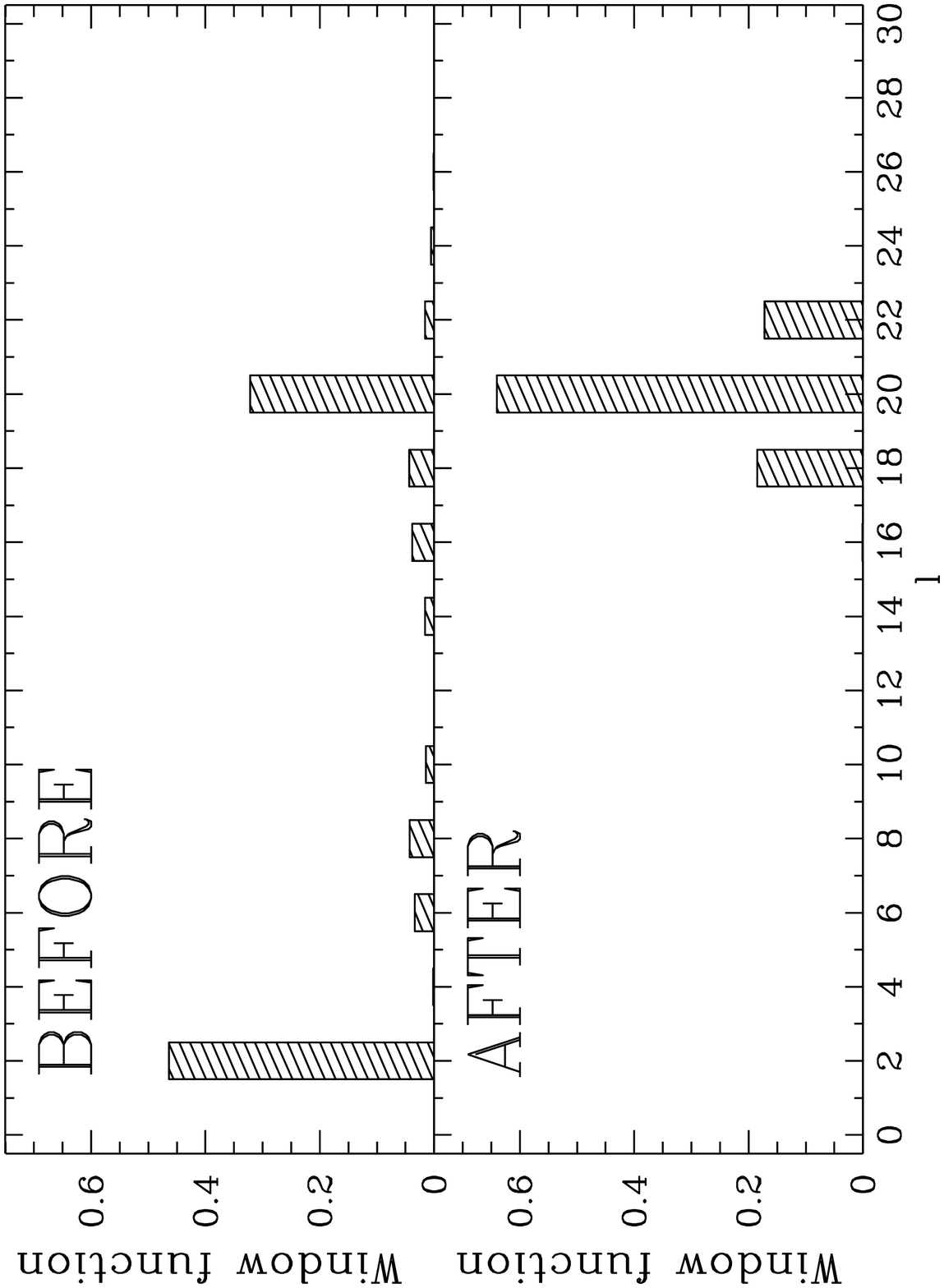}}}}
\caption{
Window functions before and after optimization.
}
Two window functions for estimation of the multipole
$\l=20$, $m=0$ are shown. The upper one is that of the
spherical harmonic method (W96), which exhibits a strong leakage
from lower multipoles such as the quadrupole.
The lower one is the one resulting from the optimal method.
\label{HistFig}
\end{figure}

\begin{figure}[phbt]
\centerline{\epsfxsize=12cm\epsfbox{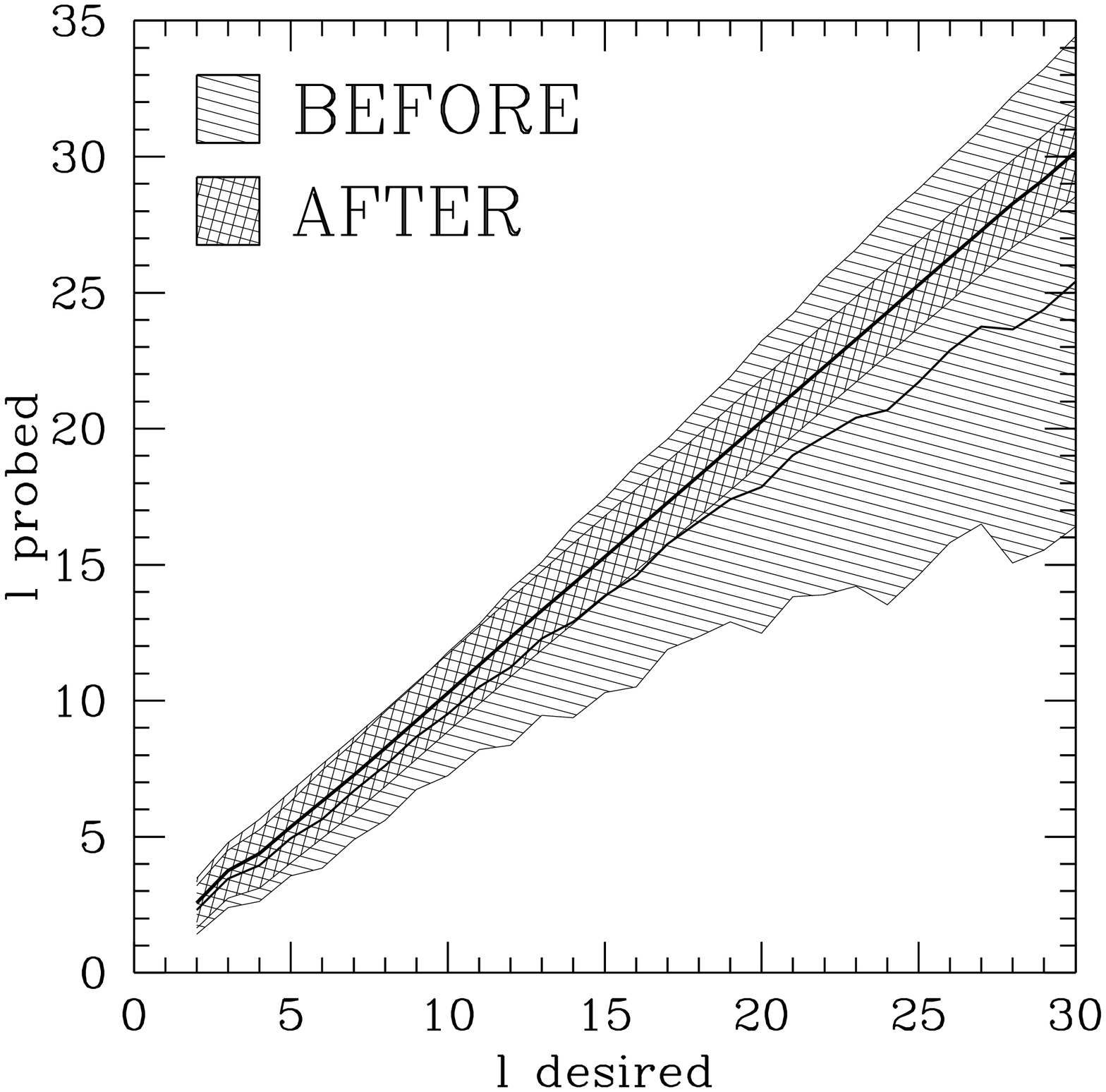}}
\caption{
Horizontal error bars before and after optimization.
}
The shaded region shows the range of $\l$-values
$\expec{\l}\pm\Delta\l$ probed by the window function devised
to estimate $D_\l$, the heavy line showing the mean $\expec{\l}$.
The wide, jagged region is the result of using spherical harmonics,
whereas the narrow strip results from the optimal method described.
\label{ComparisonFig}
\end{figure}

A number of other linear techniques for CMB analysis have recently been
published and applied to the COBE data.
Both the Karhunen-Lo\`eve (KL) signal-to-noise eigenmode method 
(B95, BS95)
and the orthogonalized spherical harmonics method
(G94) are devised to solve a different problem than that 
addressed here. 
If one is willing to parametrize the power spectrum by a small
number of parameters, for instance a spectral index and an amplitude,
then these methods provide an efficient way of estimating 
these parameters via a likelihood analysis. 
Why cannot the basis functions of these methods be used to 
estimate the angular power spectrum $C_\l$ directly, as
they are after all orthogonal over the galaxy-cut sky?
The answer is that these basis functions are orthogonal to 
{\it each other}, whereas in our context, we want them to be 
as orthogonal as possible not to each other but to the 
{\it spherical harmonics}. This is illustrated in 
\fig{HistFig}, which contrasts $\l^*=20$ window functions
of the optimal method and the generalized Hauser-Peebles method
(W94, de Oliveira-Costa \& Smoot 1995).
We want the window function to be centered on $\l^*=20$, and
be as narrow as possible, so the lower one is clearly preferable.
\Eq{WindowEq} shows that the window function will vanish for 
a given $\l$ if the weight function is orthogonal to all spherical
harmonics with that $\l$-value, so we can interpret \fig{HistFig}
as the optimal weight function being essentially 
orthogonal to all spherical harmonics except $\l=18$, 
$\l=20$ and $\l=22$ (the reason that $\l=19$ and $\l=21$ do 
not cause a problem is that even and odd multipoles remain orthogonal 
after the galactic cut, since parity symmetry is preserved).
The other weight function is
seen to couple strongly to many of the lower multipoles, 
and picks up a contribution from the quadrupole that is even 
greater than that from $\l=20$. This of course renders it quite inappropriate
for estimating the power at $\l=20$.
Analogous window functions can of course be computed for 
the orthogonalized spherical harmonics of G94 or 
the signal-to-noise eigenmodes of B95 and BS95, and 
they also exhibit window functions
that are broader than the optimal one in \fig{HistFig} --- our 
derivation of the optimal weight functions of course guarantees that
{\it any} other basis functions will give broader window functions.
However, it should be emphasized that generating such window functions
for the basis functions of G94, B95 and BS95 would be quite an 
unfair criticism of these methods, since this would be grading them 
with respect to a property 
that they were not designed to have. 
These authors have never claimed that their basis functions were optimal
for multipole estimation, merely (and rightly so) that they were 
virtually optimal for parameter fitting with a likelihood analysis.

The data points in \fig{PowerFig} are
placed at the mean $\l$-values obtained from histograms like
that in \fig{HistFig}, and as mentioned, the horizontal error bars are 
simply the r.m.s. widths of the histograms about this $\l$-value.
The horizontal error bars resulting from the optimal method are contrasted
with those obtained by the generalized Hauser-Peebles method 
in \fig{ComparisonFig}. It is seen that the former gives error bars 
$\Dl\approx 1$, whereas the latter gives a resolution 
$\Dl$ that increase strongly with 
$\l$. The new method is seen to more than double the spectral resolution
at $\l=15$ and more than triple it at $\l=20$. 
At $\l=300$, the gain would be about a factor of 100 
for a high-resolution all-sky map with a $20^\circ$ galactic cut.

As the low multipoles $C_\l$ are intrinsically much larger,
histograms like that of \fig{HistFig} exhibit mainly 
leakage
from the left, from lower multipoles. 
Thus the generalized Hauser-Peebles method probes
an effective $\l$-value $\expec{\l} < \l$ for large $\l$, as
pointed out in W94. It is seen that the new method eliminates this 
``read leak" problem.

There is quite an interesting connection between this method and the 
KL signal-to-noise eigenmode technique
of B95 and BS95. The KL-method is that which optimizes the signal-to-noise 
ratio, {\ie}, loosely speaking that which minimizes the
vertical error bars without regard for the horizontal
ones. Indeed, by taking $\scale\to\infty$
(which corresponds to dropping the first term in 
\eq{QdefEq}) and choosing $\mu_\l$ to be the expected 
power spectrum, it is readily seen that the present method reduces
to the KL method.
In other words, the KL method is a special case of the method
presented here, corresponding to ignoring the 
spectral resolution.

\subsection{Qualitative features of the method}

\fig{ComparisonFig} can be easily understood from the analogy 
with quantum mechanics discussed in the previous section.
Consider a free particle constrained to the surface of a sphere. 
If its wavefunction is required to vanish in some region, then it cannot
be in an angular momentum eigenstate, and the Heisenberg 
uncertainty principle tells us that $\Delta\l$ must be 
bounded from below by some positive constant
$\Delta\l_*$. Which wavefunction minimizes $\Delta\l$ 
given this constraint?
Basically, $\vek$ does.
So when the forbidden
region is a $20^{\circ}$ strip above and below the equator, 
$\Delta\l_*$ corresponds to half the height of the double-shaded strip.
If the wave-function is abruptly truncated at the edge of this strip
(as is the case if we simply use spherical harmonics or orthogonalized
versions thereof),
this sharp edge causes considerable ringing in Fourier
space, which is why the single-shaded region is so much wider.
Like the Gaussian minimum-uncertainty states, the optimal 
set of orthogonal functions $\vek$ fall off smoothly rather than abruptly as they
approach the galactic cut. 
This difference is readily seen in \fig{MapFig}.
Although the difference between the naive and optimal 
weight functions may appear minor in real space (\fig{MapFig}),
they are seen to be quite radical in the Fourier (multipole) 
domain (\fig{HistFig}).

\subsection{Implications for future experiments}

The fact that our method produces a spectral resolution
$\Dl$ of order $1/\Delta\theta$, independent of $\l$, 
where $\Delta\theta$ is the smallest survey dimension, 
has quite encouraging implications
for future CMB experiments.
It means that a high-resolution 
satellite experiment covering a square 
radian on the sky can produce $\Dl\approx 1$ all the way
out to the Doppler peaks, thus allowing very precise determination
of cosmological parameters.
If a long-duration balloon flight (or a satellite map with
dusty regions removed) produces a map of a 
$10^{\circ} \times 10^{\circ}$ patch of sky, then
$\Dl\approx 6$ and the location of a Doppler peak at
say $\l=200$ could be pinpointed to an accuracy 
$\Dl/\l$ of a few percent. 
When designing such future CMB experiments, it should be borne in mind
that since the {\it smallest} dimension is 
what limits the resolution, a square map is always superior 
to a rectangular one of the same area.

\bigskip
The author wishes to thank Ted Bunn, George
Efstathiou, Carlos Frenk, Joseph Silk, George Smoot and
an anonymous referee for useful comments on the manuscript.
This work was partially supported by European Union contract
CHRX-CT93-0120 and Deutsche Forschungsgemeinschaft
grant SFB-375.
The COBE data sets were developed by the NASA
Goddard Space Flight Center under the guidance of the COBE Science
Working Group and were provided by the NSSDC.


\clearpage

\section{References}

\rf Bond, J. R. {\etal} 1994;Phys. Rev. Lett.;72;13

\rg Bond, J. R. 1995;Phys. Rev. Lett.;74;4369;B95
     
\rg Bunn, E.F. \& Sugiyama, N. 1995;ApJ;446;49;BS95

\rf de Oliveira Costa, A. \& Smoot, G.F. 1995;ApJ;448;477 
 
\rg G\'orski, K. M. 1994;ApJ;430;L85;G94

\rf G\'orski, K. M. {\etal} 1995;ApJ;444;L65 

     
\rf Hauser, M. G. \& Peebles, P. J. E. 1973;ApJ;185;757

\rf Hu, W. \& Sugiyama, N. 1995;Phys. Rev. D;51;2599 




\rf Lineweaver, C. H. {\etal} 1994;ApJ;436;452

\rf Peebles, P. J. E. 1973;ApJ;185;413

\rf Smoot, G.F. {\etal} 1992;ApJ;396;L1

\rf Stompor, R., G\'orski, K. M. \& Banday, A. J. 1995;MNRAS;277;1225

\rg Tegmark, M. \& Bunn, E. F. 1995;ApJ;455;1;TB95

\rg Tegmark, M. 1995;ApJ;455;429;T95

\rg White, M. \& Bunn, E. F. 1995;ApJ;450;477;WB95
     
\rf White, M., Scott, D. and Silk, J. 1994;ARA\&A;32;319

\rf Wright, E.L. {\etal} 1994a;ApJ;420;1

\rg Wright, E.L. {\etal} 1994b;ApJ;436;443;W94


\newpage

\newpage

\newpage

\newpage

\end{document}